\documentstyle[12pt]{article}
\begin{document}
%\draft

\begin{flushright}
TIFR-TH-94/49
\end{flushright}
\begin{center}
\vspace{3 ex}
{\Large\bf
GEOMETRIC ENTROPY OF}\\
\vspace{1 ex}
{\Large\bf NON-RELATIVISTIC FERMIONS}\\
\vspace{1 ex}
{\Large\bf AND TWO DIMENSIONAL STRINGS}\\
\vspace{8 ex}
Sumit R. Das \\
Tata Institute of Fundamental Research \\
Homi Bhabha Road, Bombay 400 005, INDIA \\
\vspace{15 ex}
\bf Abstract\\
\end{center}
\vspace{2 ex}
We consider the geometric entropy of free nonrelativistic fermions in
two dimensions and show that it is ultraviolet finite for finite fermi
energies, but divergent in the infrared. In terms of the corresponding
collective field theory this is a {\em nonperturbative} effect and is
related to the soft behaviour of the usual thermodynamic entropy at
high temperatures. We then show that thermodynamic entropy of the
singlet sector of the one dimensional matrix model at high
temperatures is governed by nonperturbative effects of the underlying
string theory. In the high temperature limit the ``exact'' expression
for the entropy is regular but leads to a negative specific heat, thus
implying an instability. We speculate that in a properly defined two
dimensional string theory, the thermodynamic entropy could approach a
constant at high temperatures and lead to a geometric entropy which is
finite in the ultraviolet.
\thispagestyle{empty}
\newpage

\def\ben{\begin{equation}}
\def\een{\end{equation}}
\def\bea{\begin{eqnarray}}
\def\eea{\end{eqnarray}}
\def\nn{\nonumber}

\def\pt{\partial_t}
\def\px{\partial_x}
\def\sqf{{\sqrt{F}}}
\def\intpm{\int_{-\infty}^{\infty}}
\def\intp{\int_0^\infty}
\def\intm{\int_{-\infty}^0}
\def\fr{f_R}
\def\fl{f_L}
\def\sqf{{\sqrt{F}}}
\def\intpm{\int_{-\infty}^{\infty}}
\def\intp{\int_0^\infty}
\def\intm{\int_{-\infty}^0}
\def\fr{f_R}
\def\fl{f_L}
\def\bd{{b^+}}
\def\dd{{d^+}}
\def\pd{{\psi^+}}
\def\tb{{\tilde b}}
\def\psib{{\bar \psi}}
\def\sqx{{\sqrt {x}}}
\def\sqmx{{\sqrt {-x}}}
\def\half{{1 \over 2}}
\def\chib{{\bar \chi}}
\def\kf{k_F}
\def\klog{{\rm log}}
\def\fbr{{\bar f_R}}
\def\fbl{{\bar f_L}}
\def\a{\alpha}
\def\onen{{1 \over {\sqrt{2N}}}}
\def\onelog{{1 \over {\sqrt{\klog N}}}}
\def\onenlog{{1 \over {\sqrt{ 2N \klog N}}}}
\def\lnn{{{\rm log} N \over 2}}
\def\lla{{1 \over 2}{\rm log}({L\Lambda \over \pi})}
\def\faa{f_{3\over 2}}
\def\fab{f_{1\over 2}}
\def\sbetav{{\sqrt{\beta}} N_F \over L}
\def\spi{{\sqrt{\pi}}}
\def\bchi{{\bar{\chi}}}
\def\sqmuf{{\sqrt{\mu_F}}}
\def\vk{\vert k \vert}
\def\bpsi{{\hat \psi}}
\def\bpd{{\hat \psi}^+}
\def\bb{{\hat b}}
\def\ba{{\hat a}}
\def\bab{{\bar b}}
\def\chid{{\chi^+}}
\def\ww{\omega}
\def\psib{{\bar \psi}}
\def\kexp{{\rm exp}}
\def\kcosh{{\rm cosh}}
\def\ksinh{{\rm sinh}}
\def\cd{{\cal D}}
\def\ktanh{{\rm tanh}}
\def\ksech{{\rm sech}}
\def\cz{{\cal Z}}
\def\ptau{\partial_\tau}

\def\pr{{\em Phys. Rev.~}}
\def\prl{{\em Phys. Rev. Lett.~}}
\def\np{{\em Nucl. Phys.~}}
\def\pl{{\em Phys. Lett. ~}}
\def\cmp{{\em Comm. Math. Phys.~}}
\def\mpla{{\em Mod. Phys. Lett.~}}

Recently the entropy of entanglement between different regions of space
in quantum field theories have been intensively studied
\cite{SORKIN}-\cite{LARWILC}. The motivation
for this is its direct connection to the question of information loss
due to black holes and black hole entropy \cite{BEK}-\cite{MATHUR}.
A significant feature of this entanglement entropy,
or ``geometric entropy'' is that it is ultraviolet divergent in typical
field theories. This has been interpreted to imply that at least at the
semiclassical level information loss due to the formation of a horizon
is inevitable in quantum field theories. The divergence of the entropy
is a reflection of short distance singularities in quantum field theories.
Alternatively \cite{BARB}-\cite{ODIN}
the divergence is related
to the behaviour of the usual thermodynamic entropy at high temperatures
since, as we shall see below, the geometric entropy effectively
involves an {\em integral} of the {\em thermodynamic entropy density}
over all temperatures.

One may hope that in string theories this divergence disppears because
of a soft ultraviolet behaviour
\cite{SUSS}. However, to leading order in the
string perturbation expansion, the thermodynamic free energy of a
string is equal to the sum of the free energies of the physical modes
of the string and one would obtain the same divergence in each term of
the sum. Furthermore, unlike in a field theory of a finite number of
fields, the thermodynamic free energy of free strings is itself
divergent at the Hagedorn temperature. This is an {\em infrared}
diveregence and might signal an {\em instability} of the theory
and one might obtain a
finite answer once one takes into account interactions and shift to a
stable vacuum \cite{ATWIT}. As argued in \cite{BARB,DABH1} the geometric
entropy in string theory is afflicted by this Hagedorn transition
and what appears as an ultraviolet divergence in each term of the sum
over all string modes may be interpreted as an infrared problem in
the full answer.

The need to include string interactions calls for a formulation of the
problem in some well defined and tractable string field theory.  While
this appears to be an almost impossible task at present, there is one
string theory where a tractable nonperturbative formulation exists, at
least for some bulk quantities.  This is the two dimensional string defined
via the one dimensional matrix model \cite{CEQONE}.  The singlet
sector of the matrix model
\footnote{ As we will see soon the nonsinglet contributions are
irrelevant for a calculation of the geometric entropy in the
ground state }
may be written as a two dimensional
collective field theory of the density variable.The fluctuations of
the collective field represent a massless particle which is the only
propagating degree of freedom of the two dimensional string
\footnote{The relationship between the collective field and
the massless scalar of the effective field theory is rather
subtle and not completely clear at this moment.
See \cite{POLC,LOUG} for a recent discussion.} and
the coupling is proportional to the inverse of the fermi energy.
For nonperturbative considerations it is better to write the model
as a field theory of nonrelativistic fermions
in the presence of an inverted harmonic
oscillator potential and no self-interactions.
The idea, then, is to consider a geometric entropy in
this model and use exact nonperturbative answers to understand stringy
effects. Hopefully this will teach us something about higher
dimensional string theories as well.

In this paper we take the first step in this program. We first consider
the problem of {\em free} nonrelativistic fermions in two
dimensions. By constructing the explicit expression for the ground
state wave functional and the corresponding geometric density matrix
we argue that the geometric entropy has no ultraviolet divergence for
finite fermi energies. In terms of the collective field theory this is
a {\em nonperturbative} phenomenon.  In fact the result follows from
the softer behaviour
of the ordinary thermodynamic entropy at high temperatures (compared
to relativistic fermions). We then
consider the the high temperature limit of the ``exact'' expression
for the thermodynamic partition function of the singlet sector
 of the one
dimensional matrix model \cite{GKLEB}. We will show that the genus
expansion breaks down at high enough temperatures and hence the
geometric entropy of this string theory is essentially
nonperturbative.  A naive high temperature expansion leads to a
regular behaviour of the entropy, but the specific heat turns out to
be {\em negative}, signifying a nonperturbative instability.

Consider a theory of fields $\phi (x,t)$ in $1 + 1$ dimensions. The
ground state wave functional may be written as
$\Psi_0 [\phi_L,\phi_R]$,
where $\phi_L ~(\phi_R)$ denotes the field $\phi$ for $x < 0 ~(x > 0)$.
The density matrix which gives expectation values of operators localized
in the $x > 0$ region is
\ben
\rho(\phi_R, \phi_R ') = \int {\cal D}\phi_L \Psi_0 [\phi_L,\phi_R]~
\Psi_0 [\phi_L,\phi_R ']
\label{eq:ten}
\een
As shown in \cite{CALWILC,KABAT}
the quantity ${\rm Tr}\rho^n$ may be represented
as an euclidean path integral over a cone with a deficit angle
$2\pi(1-n)$. The geometric entropy may be then written using a
``replica trick'' as
\ben
S_g = -{\hat \rho}\klog~{\hat \rho} = [(1 - n{d \over dn})\klog
{\rm Tr}~\rho^n]_{n=1}
\label{eq:eleven}
\een
For relativistic systems this establishes the equality of the geometric
entropy with the usual {\em thermodynamic} entropy of the field in
Rindler space at the Rindler temperature, and hence the quantum correction
to the entropy of a large mass black hole.
For nonrelativistic systems
the Rindler hamiltonian would depend on the Rindler time.  However
the geometric entropy defined above makes sense and ${\rm Tr}\rho^n$
is still a path integral on a cone.

Consider a system of $N_F$ free nonrelativistic fermions
contained in a box : $-L < x < L$. The dispersion relation for
the single particle states is given by $e (k) = \half k^2$ where
$k$ is the (spatial) momentum which is quantized as $k = {\pi n
\over L}$ with integer or zero $n$. However, below we will
often use continuum notation.
The second quantized fermion field operator can be expanded in
terms of quasiparticle operators $\bb (k)$ and $\bb^+ (k)$
\ben
\bpsi(x,t)  =  \intpm {dk \over 2\pi}e^{-ikx} [\bb (k)
\theta(\vk - k_F) + \bb^+(-k) \theta(k_F - \vk)]
\label{eq:qone}
\een
and similarly for $\bpsi^+$.
The operators $\bb, \bb^+$ satisfy the standard anticommutation
relations $\{ \bb(k,t), \bb^+(k',t) \} = \delta (k-k')$.
The ground state is then described by the filled Fermi sea
$\bb(k) |0> = 0$

In terms of the coherent states of the $b$-oscillators
\ben
\bb(k) |b(k) > = b(k) |b(k)>~~~~~~~~<b(k)| \bb^+ (k) = \bab(k) <b(k)|
\label{eq:qfour}
\een
where $b(k),\bab (k)$ are grassman numbers, the ground state wave
functional is given by
\ben
\Psi_0[b(k),\bab(k)] = exp[-\half\intpm {dk \over 2\pi} \bab (k) b(k)]
\label{eq:qfive}
\een
The grassmann fields appearing in the path integral are, however, not
$b(k)$ and $\bab (k)$. Rather, they are the combinations
\ben
\psi (k) = b(k) \theta (\vk - k_F) + \bab (-k) \theta (k_F - \vk)
\label{eq:fivea}
\een
and similarly for $\psib (k)$. The fourier transforms of these fields
are the original fields $\psi (x)$.
For reasons which will be clear in a moment we will use a new
field $\chi (q) = \psi (k_F + q)$. Then the wavefunctional (\ref{eq:qfive})
becomes
\ben
\Psi_0 = {\rm exp}~[-\half(\intpm
{dq \over 2\pi} \chib (q) \chi (q) - 2 \int_{-2k_F}^0
{dq \over 2\pi} \chib (q) \chi (q))]
\label{eq:qeight}
\een
The inverse fourier transform of $\chi(q)$,
denoted by $\chi (x)$ is related to the original field $\psi (x)$
by $\chi(x) = e^{ik_Fx}\psi (x)$. Since this relationship is
{\em local} we can compute the geometric entropy using these fields.

In (\ref{eq:qeight}) the first term may be written as a integral
over position space of a {\em local} quantity $\chib (x) \chi (x)$
and do not contribute to the geometric entropy. The only momentum
modes which contribute to the geometric entropy are those in the
filled Fermi sea as in the second term in (\ref{eq:qeight}). The
fermi momentum thus roughly acts as an ultraviolet cutoff and this
is the essential reason why the geometric entropy turns out to be finite.

We can proceed similarly for a relativistic Weyl fermion. For example
for right moving fermions the fermi momentum is zero and the Dirac sea
consists of all negative momenta. It is easy to see that the
wavefunctional is the expression (\ref{eq:qeight}) where the limit of
integration $-2k_F$ in the second term is replaced by $-\infty$.
Clearly, in the limit of large $k_F$ the nonrelativistic expression
reduces to the relativistic expression. This simply reflects the fact
that the excitations very close to the fermi level behave as massless
relativistic particles with the velocity of light replaced by the
fermi velocity.

To compute the geometric entropy we now introduce the following
technique which may be trivially generalized to other situations
(e.g. relativistic bosons, fermions etc.). We first expand the
field eigenvalues in terms of modes which are localized to the
left region ($x < 0$), $\fl (\ww)$ and those localized to the
right region ($x > 0$), $\fr (\ww)$ as follows
\ben
\chi (x)  =  {1 \over {\sqrt{| x |}}}[\theta (x) \intpm d \ww
( {x \over a} )^{-i\ww} \fr (\ww)
+ \theta (-x) \intpm d \ww
( {-x \over a} )^{-i\ww} \fl (\ww)]
\label{eq:qten}
\een
where $a$ denotes the lattice spacing (or some other ultraviolet
cutoff).
The factor of ${1 \over {\sqrt{| x |}}}$
arises from the dimension of the field $\chi (x)$ under rescalings
of $x$.
%\footnote{ In
%our nonrelativistic theory we assign the following dimensions :
%$dim~[t] = -1$, $dim~[x] = -\half$. Then $dim~[\psi] = {1 \over 4}$}.
To obtain the expression for $\chid (x)$ replace $\fl,\fr$ by
$\fbl, \fbr$ and $\ww$ by $-\ww$ in the integrand of (\ref{eq:qten}).

The fields $\chi (q)$ and $\chi^+(q)$ may be now expressed in terms
of $\fl$ and $\fr$. We get the following expressions, all valid for
$q > 0$.
\bea
\chi (q) & = & \intpm {d\ww \over 2\pi}[ie^{\pi\ww}\fl(\ww)
+\fr(\ww)]G(q,\ww) \nonumber \\
\chib (q) & = & \intpm {d\ww \over 2\pi}[\fbl (\ww)
+ ie^{-\pi\ww} \fbr (\ww)] G(q,-\ww) \nonumber \\
\chi (-q) & = & \intpm {d\ww \over 2\pi}[\fl (\ww)
+ ie^{\pi\ww} \fr (\ww)] G(q,\ww) \nonumber \\
\chib (-q) & = & \intpm {d\ww \over 2\pi}[ie^{-\pi\ww}\fbl (\ww)
+ \fbr (\ww)] G(q,-\ww)
\label{eq:qfifteen}
\eea
where we have defined
\ben
G(q,\ww) = \intp {dx \over {\sqrt{x}}}
 e^{iqx}({x \over a})^{-i\ww} = a^\half ({i\over q a})^{\half -
i\ww}~\Gamma (\half - i\ww)~~~~~~~~q > 0
\label{eq:qtwelve}
\een
Note that $G(-q,\ww) = i~e^{\pi\ww}~G(q,\ww)$,
which has been used in writing (\ref{eq:qfifteen}).

The ground state wave functional may be now rewritten explicitly
in terms of $\fl , \fr$ :
\ben
\Psi_0[\fl,\fr]=\kexp~[-\half(I(\infty)+{\bar I}(\infty)-2{\bar I}(2k_F))]
\label{eq:qnineteen}
\een
where we have defined
\ben
I(\gamma)  \equiv \int_0^\gamma {dq \over 2 \pi}
\chib (q) \chi (q);~~~~~~~~
{\bar I}(\gamma) \equiv \int_{-\gamma}^0 {dq \over 2 \pi}
\chib (q) \chi (q) =
\int_0^\gamma {dq \over 2 \pi}\chib (-q) \chi (-q)
\label{eq:qsixteen}
\een
In evaluating $I(\gamma)$ we need to perform the integral
\ben
F_\gamma (\ww , \ww') = \int_0^\gamma {dq \over 2\pi} G(q,\ww)~G(q,-\ww ')
\label{eq:qeighteen}
\een
In the presence of a finite box size $2L$ the lower limit of the $q$
integral is really ${\pi \over L}$.
Using (\ref{eq:qfifteen}) one may write the wave functional
explicitly in terms of $\fl$ and $\fr$ and obtain
the density matrix for fields localized in the $x > 0$ region
by simply functionally integrating over $\fl ,\fbl$
\ben
\rho(\fr, \fr ')=\int \cd \fbl \cd \fl~\Psi_0[\fr,\fl]~\Psi_0[\fr ',\fl]
\label{eq:qtwenty}
\een

It is easy to evaluate the integral $F_\gamma(\ww, \ww')$ for
$\gamma = \infty$. The result is
\ben
F_\infty(\ww , \ww ') = {2\pi i}~\ksech(\pi\ww)
\delta (\ww - \ww ')
\label{eq:qtwoone}
\een
{}From this we can make the first consistency check on our technique.
Using (\ref{eq:qtwoone}) one
gets the wave functional for relativistic fermions
\bea
\Psi_0^{rel} & = \kexp~[-{2\pi}\int d\ww~
& [\ktanh~\pi \ww(\fbr(\ww)\fr(\ww)-\fbl(\ww)\fl(\ww)) \nn \\
 & & + i ~\ksech~\pi \ww(\fbl(\ww)\fr(\ww) -\fbr (\ww) \fl (\ww))]
\label{eq:qtwotwo}
\eea
After a trivial rescaling of the fields by a constant
\footnote{It may be easily verified
from the definitions that a rescaling of fields by a constant does not
change the geometric entropy, as one would expect.}
this is exactly the answer
obtained by euclidean path integral methods in \cite{LARWILC}.
As discussed above, (\ref{eq:qtwotwo}) is also the answer for
nonrelativistic fermions exactly at $k_F = \infty$.

The density matrix and its various powers in this relativistic limit
may be obtained as in \cite{LARWILC} and the geometric entropy may
be calculated using the replica trick to obtain the result
\ben
S_g \sim \klog (L/a)
\label{eq:qnine}
\een
This expression has an ultraviolet as well as an infrared divergence.
This result may be understood in a very simple way.
The geometric entropy is a measure of the entanglement of the
modes $\fl$ and $\fr$.
It is clear from the wave functional that the modes $\fl (\ww)$ and
$\fr (\ww)$ mix for {\em all} values of $\ww$. This means that all the
$\ww$-modes contribute to the geometric entropy, which should be
then proportional to the total number of $\ww$'s.
{}From (\ref{eq:qten}) it follows that for large $L$
the number of allowed $\ww$'s is proportional to $\klog(L/a)$ -
hence the answer (\ref{eq:qnine}).

For small values of $k_F$ the integral ${\bar I}(2k_F)$ may be
approximated as
\ben
{\bar I}(2k_F) \sim (2k_F - {\pi \over L})~\chib (0) \chi(0)
\label{eq:sone}
\een
It may be easily seen that for
large $N = {L \over a}$ the function $G(0,\ww)$ is peaked at
$\ww = 0$, falls to zero at $\ww \sim {1 \over \klog N}$ and
then oscillates rapidly around zero. This means that the mode
$\chi (0)$ expressed as an integral over $\ww$ essentially
receives contribution from a few values of $\ww$ around $\ww =0$.
To the lowest order we can then approximate the wave functional
by
\bea
\Psi_0[\fl,\fr] & \sim \kexp &[- \intpm d\ww
(\fbl(\ww)\fl(\ww) + \fbr(\ww) \fr(\ww)) \nn \\
& & + 2\klog~({k_F L \over \pi})~[-\fbr(0)\fr(0)
- \fbl(0)\fl(0) \nn \\
& & +i\fbr (0)\fl(0) - i\fbl(0)\fl(0)]
\label{eq:twofour}
\eea
In (\ref{eq:twofour}) the original integrals over $\ww$ have been
replaced by sums and the resulting factors of $\klog N$ have been
suitably absorbed by a redefinition of the fields.

In (\ref{eq:twofour}) the left and right modes mix only for $\ww =
0$. For small but finite $k_F$ only a few modes around $\ww = 0$ mix,
and these alone
contribute to the geometric entropy.  Let us evaluate the entropy due
to the $\ww = 0$ mode alone. The final result will be this
contribution multiplied by a factor of order unity.  The unnormalized
density matrix for this mode is easily seen to be
\bea
\rho(\fr, \fr ') & = \kexp[- & (1-\half\ksech^2 \eta)(\fbr(0)\fr(0)+
\fbr '(0) \fr '(0)) \nn \\
 & & + \half \ksech^2 \eta(\fbr (0) \fr'(0) + \fbr' (0) \fr (0))]
\label{eq:twofive}
\eea
where we have defined
\ben
\ksech~\eta = {\klog{k_F L \over \pi} \over
\klog{k_F L \over \pi} + 2}
\label{eq:twosix}
\een
By a redefinition of the fields $\fr \rightarrow (2{\rm tanh}
{}~\eta)^\half~\fr$ one may rewrite the density matrix in a form
which facilitates the calculation of ${\rm Tr}\rho^n$ for any $n$.
The final result for the geometric entropy is
\ben
S_g = 2\klog(2\kcosh~\eta) - 2\eta~({\rm tanh}~\eta)
\label{eq:twoeight}
\een
For $k_F L \sim \pi$ one has $\eta \rightarrow \pm \infty$ and
the leading order expression for the geometric
entropy follows from (\ref{eq:twoeight})
\ben
S_g^{(0)} = (\klog{k_F L \over \pi})^2 [1 + \klog~2 -
2~\klog~(\klog{k_F L \over \pi})]
\label{eq:twonine}
\een
The total geometric entropy for small $k_F$
is $S_g^{(0)}$ multiplied by a factor of order unity.
As advertized above the result does not involve the ultraviolet
cutoff essentially because only a few modes contribute to the
entropy, rather than all of the $\klog~N$ modes. In a similar
fashion we expect that for finite $k_F$ the role of the ultraviolet
cutoff is replaced by $k_F$.

There is an alternative way to view the quantity ${\rm Tr}\rho^n$.
Consider dividing up the region $x > 0$ into small cells. In the
thermodynamic limit the path integral which represents ${\rm
Tr}\rho^n$ would be a product of path integrals for each individual
cell.  However the path integral for the cell centered at the point
$x$ is the standard {\em thermodynamic} partition function at a
temperature $T(x) = {1 \over 2\pi n x}$. Thus the geometric
entropy may be obtained by simply calculating the
thermodynamic entropy {\em density} at a temperature $T(x)$ and
integrating the result from $x=0$ to $x = \infty$. This is the
procedure used to compute the genus one contribution to the
entropy of strings in \cite{BARB,DABH1}.

The ultraviolet divergence of the geometric entropy in relativistic
field theories may be now understood in terms of the behaviour of
the standard thermodynamic entropy at high and low temperatures. For
example, a free massless boson in $d$ space
dimensions at temperature $T = {1 \over \beta}$ has an entropy density
$s = {A \over \beta^d}$. This diverges for
low $\beta$, or high temperature for all $d$.
In our problem we have to put
$\beta = 2\pi x$ and integrate this entropy density over $x$, so that
there is a diveregence of the geometric entropy from the lower limit
of integration. This is the ultraviolet divergence in the entropy.
In the corresponding Rindler problem, the divergence arises because
the local temperature is very high near the horizon and the corresponding
contribution to the entropy density is large. For $d = 1$
there is an additional infrared divergence coming from large $x$. Thus
using an ultraviolet cutoff $\epsilon$ and an infrared cutoff $L$ in the
integral over $x$ one has
$S \sim  {1 \over \epsilon^{d-1}}$ for $d > 1$ and
$S \sim \klog({L \over \epsilon})$ for $d = 1$.

Consider now our system of $N_F$ free nonrelativistic fermions in one
spatial dimension in the grand canonical ensemble with
an inverse temperature $\beta = {1 \over T}$ and fugacity $z$.
The fugacity is determined in terms of $N_F$ by $N_F = z{\partial_z}~
\klog~\cz$ (where $\cz$ is the partition function) which leads to
\ben
N_F = {L \sqrt{2 \over \pi \beta}} \fab (z) ~~~~~\fab (z) = z\partial_z
\faa (z)
\label{eq:nineteen}
\een
where the function $\faa$ is defined as
\ben
\faa (z) = {1 \over \spi}\intpm dx~\klog~(1 + z~e^{-x^2})
\label{eq:sixteen}
\een
The expression for the entropy follows from usual thermodynamics and
is given by
\ben
S = N_F({3 \faa (z)\over 2 \fab (z)} - \klog~z)
\label{eq:twenty}
\een

The function $\fab (z)$ is a monotonically increasing function of $z$.
Thus for large values of $\sbetav$
(i.e. low temperatures or high densities) one may
use the standard expansions of the functions $f_r (z)$ for
large $z$ \cite{HUANG}
to obtain the leading order term in the entropy density
\ben
s = {N_F \over L}[{\pi^2 \over 6 \klog~z}
+ {13\pi^4 \over 360 (\klog~z)^3} +
\cdots]  =
{\pi \over 6 \beta {\sqrt{2\epsilon_F}}} + {8 \pi^3 \over
45~\beta^3~k_F^5} + \cdots
\label{eq:twentythree}
\een
The leading term is the same expression for a free relativitsic
boson or a free relativistic fermion in two dimensions,
apart from a factor of ${\sqrt{2\epsilon_F}}$. The reason
is simple. For a given temperature and size of the system,
large values of $\sbetav$ mean
large fermi momenta. In this case the excitations are restricted
to particle-hole excitations near the fermi level. The dispersion
for these excitations are given precisely by $E(k) = {\sqrt{2\epsilon_F}}
\vert k \vert$ which is the same as that of a massless relativistic boson
in two dimensions. The factor of ${\sqrt{2\epsilon_F}}$ in the
dispersion relation explains the same factor in the entropy
density. What is more
significant for our considerations is the fact that the same result
applies for a given $\mu_F$ and size, but low temperatures.

On the other hand, for small values of $\sbetav$ one has to use the
power series expansions of $f_r (z)$ for small $z$ \cite{HUANG}
and one easily gets for the entropy density
\ben
s = {N_F \over L} - {N_F \over 2L}\klog({2\pi \beta N_F^2\over L^2}) +
O(\beta^\half)
\label{eq:twentyfour}
\een
To obtain the geometric entropy one has to integrate $s(2\pi x)$
over all positive $x$. If we use the low temperature expansion
we would get an ultraviolet divergent answer from the behaviour
of the integrand near $x = 0$. In fact it may be easily checked
that the lowest order answer agrees completely with the direct
calculation of the geometric entropy of relativistic
fermions discussed above. However to treat the region $x = 0$
we have to use the high temperature behaviour in (\ref{eq:twentyfour}).
This shows that the integrand has an
{\em integrable} singularity at $x = 0$ whereas there is a logarithmic
divergence coming from large values of $x$. Thus the geometric entropy
is finite in the ultraviolet. This is in marked contrast with
relativistic bosons or fermions. Essentially as one approaches the
point $x = 0$ the temperature becomes large compared to the scale
set by the fermi energy. Thus the fermi energy provides a cutoff to
the relativistic behaviour for large values of $x$.

The true significance of this result can be appreciated if one
rewrites the model in terms of the collective field theory of
the density of fermions $\rho(x,t)$. Consider a more general
system of $N_F$ fermions in an external static potential $V(x)$.
The density $\rho (x,t)$ may be expanded around the ``classical''
average value
\ben
\rho (x,t) = \rho_0 (x) + \partial_x \xi (x,t) ~~~~~~
\rho_0 = {1 \over \pi}{\sqrt{2(\mu_F - V(x))}}
\label{eq:cone}
\een
where $\mu_F$ is the fermi level.
Introduce the ``time of flight'' variable
$\tau (x) = {1 \over \pi}\int^x {dx \over \rho_0(x)}$. Then the
hamiltonian which governs the fluctuations $\xi (\tau,t)$ is
given by
\ben
H_{coll} = : \half \int d\tau[\Pi_\xi^2 + (\ptau \xi)^2
-{1 \over \pi^{3 \over 2}~\rho_0^2}(\Pi_\xi (\ptau \xi) \Pi_\xi
+ {1 \over 3}(\ptau \xi)^3) - {1 \over \pi^{5 \over 2}}(
{\rho_0 '' \over 3\rho_0^3}-{(\rho_0 ')^2 \over 2 \rho_0^4})] :
\label{eq:ctwo}
\een
where $\Pi_\xi$ denotes the field momentum.
Note that the hamiltonian already comes in the normal ordered
form \cite{DAS}
and leads to finite answers. The density thus
behaves as a scalar field with in general position dependent
coupling given by ${1 \over \rho_0^2}$.
For free fermions $V(x)=0$. Then $\rho_0 = {\sqrt{2\mu_F}}$ and
$\tau = {{\sqrt{2\mu_F}}\over \pi}$.
The coupling is a constant and equal to ${1 \over 2 \mu_F}$.
For $\mu_F$ large compared to typical energies,
one has a {\em free} relativistic scalar field,
which is why in this limit we obtained the relativistic answer
for the entropy. For small $\mu_F$ the theory is strongly coupled
and the perturbation expansion does not make sense. Thus the
finiteness of the entropy demonstrated above is a {\em non-perturbative}
effect in this collective field theory.

Let us now consider the one dimensional matrix model which provides
a nonperturbative definition for the two dimensional noncritical
string. This is defined by the action
\ben
A = \lambda \int dt {\rm Tr}~(\half [\pt M(t))^2 + V(M(t))]
\label{eq:mone}
\een
where $M(t)$ is a $N \times N$ matrix and the potenial function $V(M)$
has to be chosen such that it has a quadratic maximum.The detailed
form of the potential is not important in the double scaling
limit. As is well known the singlet sector of this model may be
written exactly in terms of nonrelativistic fermions $\psi (x,t)$
where $x$ denotes the space of eignevalues of the matrix $M$. These
fermions have no self interactions, but move in an external potential
$V(x)$. Let us denote the fermi energy by $\mu_F$. Then as the
coupling $g = {N \over \lambda}$ of the theory approaches a critical
value $g_c$ the fermi level $\mu_F$ appraches $\mu_c$ which is the
energy at the top of the potential hump. The double scaling continuum
limit of this problem is given by $\mu = \mu_c - \mu_F \rightarrow 0$
and $\lambda \rightarrow \infty$ with
$\kappa = \lambda\mu = {\rm fixed}$. This is
known to describe the two dimensional string. In this limit only the
quadratic hump of the potential is relevant and one has a problem of
nonrelativistic fermions in two dimensions in an external potential
$V(x) \sim -x^2$ - the inverted harmonic oscillator. The collective
field theory hamiltonian is given by (\ref{eq:ctwo}) with this
potential. This is now a version of string field theory and the
string coupling is $g_{st} \sim {1 \over \kappa}$.

The geometric entropy of this model in the ground state may be calculated
in principle by the technique discussed above. The expansion for
the fermion fields are no longer in terms of plane waves, but in terms
of the eigenfunctions of the Schrodinger operator in the inverted
harmonic oscillator potential, which are parabolic cylinder functions.
The main modification would be to replace the plane waves $e^{iqx}$ by
parabolic cylinder functions in the definition of $G(q,\ww)$ in
(\ref{eq:qtwelve}). The resulting expressions are rather difficult
to analyze.

We have seen, however, that the divergence of the geometric entropy
is related to the high and low temperature behaviors of the ordinary
thermodynamic entropy. The complete thermodynamic partition function
of the matrix model is a formidable task since this includes
contributions from the non-singlet states. However if we are interested
in computing the geometric entropy for the {\em ground state},
which is a singlet, the
trace involved in ${\rm Tr}\rho^n$ is a trace over {\em singlet} states
alone. The thermodynamics in the singlet sector has been completely
solved and we can use the known results of \cite{GKLEB}

Let us define $\Delta = g_c - g$ and let $F$ denote the free energy.
Then the chemical potential $\mu$ is determined by the equation
\ben
{\partial \Delta \over \partial \mu}
= {1 \over 2\pi} {\rm Re}~[\int_0^\infty~{dt \over t}~e^{-it}~
{(t/\kappa )\over \ksinh (t/\kappa )}~
{(\pi t/\kappa \beta)\over \ksinh (\pi t/\kappa\beta)}]
+ {1 \over 2\pi}[ \klog {\lambda \over 2} - \gamma]
\label{eq:mtwo}
\een
The infinite constant $\gamma$ is necessary for (\ref{eq:mtwo})
to reproduce the correct WKB expansion in ${1 \over \lambda}$,
but would be unimportant in what follows.
The free energy $F$ is then obtained by integrating the equation
\ben
{\partial F \over \partial \Delta}=\lambda^2(\mu - \mu_c)
\label{eq:mthree}
\een
Recalling that the number of fermions is $N = \lambda g$ one may
write down the expression for the entropy using standard relations
in the grand canonical ensemble
\ben
S = \beta^2 [ ({\partial F \over \partial \beta})_\mu
- \lambda^2 \mu ({\partial \Delta \over \partial \beta})_\mu]
\label{eq:mthreea}
\een
The second term in (\ref{eq:mthreea}) arises because we are
considering derivatives with fixed $\mu$ rather than fixed $N$.

The above expressions have an important duality symmetry. From
(\ref{eq:mtwo}) it follows that
\ben
{\partial \Delta \over \partial \mu}(\beta,\lambda)
= {\partial \Delta \over \partial \mu}({\pi^2 \over \beta},{\lambda
\beta \over \pi})
\label{eq:mtwok}
\een
It then follows from (\ref{eq:mthree}) that the canonical partition
function $\beta F$ is {\em invariant} under this duality transformation.

The standard genus expansion is obtained by
considering $\kappa$ to be large and expanding the hyperbolic sine
functions in a power series expansion. The result for this asymptotic
expansion is
\ben
{\partial \Delta \over \partial \mu} = {1 \over 2\pi}[-\klog~\mu
+ \sum_{m=1}^\infty {f_m(\beta) \over (\beta\kappa^2)^m}]
\label{eq:mfour}
\een
where the functions $f_m(\beta)$ are symmetric under the duality
transformation ${\beta \over \pi} \rightarrow {\pi \over \beta}$
and has the form
\ben
f_m (\beta) \sim \sum_{k=0}^m~C(m,k)~(\beta)^{m-2k}
\label{eq:mfive}
\een
$C(m,k)$ are numbers related to Bernoulli coefficients.
The $m = 1$ in the sum in the above expression is the one
loop contribution in string theory. The corresponding one loop
free energy is
\ben
F = {\klog~\mu \over 12\pi\beta}({\beta \over \pi} + {\pi \over \beta})
\label{eq:msix}
\een
This result is identical to the free energy of an ideal gas of massless
bosons in one spatial dimension of size $\klog~\mu$ apart from a
constant which is in fact the {\em finite} one loop correction to
the ground state energy of the system. This is expected since the
only propagating mode of the two dimensional string is a massless
scalar. The same result is obtained by performing the
Polyakov path integral in the continuum $d = 2$ string theory
\cite{BERKLEB} and is in fact the result of the following modular
invariant integral
\ben
F = \int_{{\cal F}}{d^2 \tau \over \tau_2^2}\sum_{m,n}~
exp[-{\beta^2 \vert n-m\tau \vert^2 \over 4\pi \tau_2}]
\label{eq:msixz}
\een
where $\tau$ is the complex modular parameter on the torus and
the integration is over the fundamental domain. The term in the
sum with $m,n = 0$ is the zero temperature free energy and is
separately modular invariant. The temperature dependent term
evaluates to the second term in (\ref{eq:msix}). As noted this
is the contribution from a single massless scalar. This is
nevertheless modular invariant as it should be since it follows
from a string theory \footnote{See \cite{ODIN} for related remarks}
. Note further that unlike in higher
dimensions the one loop answer does not have a Hagedorn behavior
at any finite temperature simply because the 2d string has only one
propagating degree of freedom.

The contribution to the geometric entropy from the one loop term
alone gives the standard logarithmically divergent answer for a
free relativistic boson in two dimensions. One may regard the entire
divergence to be {\em infrared} since the answer is modular invariant,
as argued in \cite{DABH1}. However the answer is really the same as
a single boson. The string interpretation then gives a relation
between the infrared and ultraviolet cutoff in terms of the
cutoff in the modular parameter ${\rm Im} \tau$.

Returning to our description of the usual thermodynamics of the
matrix model let us record below the genus expansion for the
thermodynamic entropy.
\ben
S = -{1 \over 3\beta}[\klog~\mu + 1] + {1 \over \lambda^2}
[{1 \over 18\beta} +{7 \over 90 \beta^3}] + \cdots
\label{eq:kone}
\een

The asymptotic expansion (\ref{eq:mfour}) makes sense for small
temperatures, i.e. large $\beta$. Indeed, for large $\beta$,
${\partial \Delta \over \partial \mu}$ approaches a constant,
and the leading correction from {\em all the terms in the sum
in} (\ref{eq:mfour}) is of order ${1 \over \beta^2}$, as is
clear from (\ref{eq:mfive}). The leading term in the
free energy is then seen to be of order ${1 \over \beta^2}$ as
well and has contributions from all genus.

We have seen, however, that the ultraviolet divergence of the
geometric entropy is related to the {\em high} temperature behaviour
of the free energy. The asymptotic expansion of (\ref{eq:mtwo})
leading to the genus expansion (\ref{eq:mfour}) clearly breaks
down for high temperatures. More specifically for $\beta < {\pi
\over \kappa}$ one should not expand the hyperbolic functions
in (\ref{eq:mtwo}) in power series.
This means that for high temperatures
strong coupling effects (in the sense of string theory) become
important. Indeed the genus expansion in (\ref{eq:mfour})
contain ever increasing powers of ${1 \over \beta}$ with higher
and higher genus giving rise to higher and higher powers.
It is clearly necessary to look at the high temperature
limit {\em for arbitrary values of the string coupling} $g_{st}$.

Before considering the behaviour  at high temperatures let us
look at the behavior at low temperatures,$\beta >> {\pi \over \kappa}$,
 but independent of the genus expansion.
This means we expand the factor ${(\pi t/\kappa \beta)/
\ksinh (\pi t/\kappa\beta)}$ in (\ref{eq:mtwo})
in a power series, but {\em not} the
first factor. This gives a power series in ${1 \over \beta^2}$ for all
values of $\kappa$. A straightforward evaluation of the integrals
yield the first few terms
\bea
{\partial \Delta \over \partial \mu} & = &  -{1 \over 2\pi} {\rm Re}~
\Psi [\half(1 + i\kappa)] + {\pi \over 48\beta^2} {\rm Re}~
\Psi '' [\half(1 + i\kappa)] \nn \\
 & &  - {7 \pi^3 \over 11520 \beta^4}
\Psi''''[\half(1 + i\kappa)] + \cdots
\label{eq:mseven}
\eea
where $\Psi (x)$ denotes the digamma function. It may be checked that
if one now considers the large $\kappa$ asymptotic expansion of the
digamma functions in (\ref{eq:mseven}), the result
agrees with the large-$\beta$ limit of the genus expansion in
(\ref{eq:mfour}).

Since we have duality invariance ${\pi \over \beta} \rightarrow
{\beta \over \pi}$ we can {\em deduce} the high temperature
behaviour from the low temperature expansion (\ref{eq:mseven}).
Using (\ref{eq:mtwok}) and (\ref{eq:mseven}) one gets
\bea
{\partial \Delta \over \partial \mu} & = & -{1 \over 2\pi} {\rm Re}~
\Psi [\half(1 + i{\kappa\beta \over \pi})] + {\beta^2 \over 48\pi^3}
{\rm Re}~
\Psi '' [\half(1 + i{\kappa\beta \over \pi})] \nn \\
 & & - {7 \beta^4 \over 11520
\pi^5} \Psi''''[\half(1 + i{\kappa\beta \over \pi})] + \cdots
\label{eq:msevenz}
\eea
The expansion in (\ref{eq:msevenz}) is now valid for small $\beta$.
For very high temperatures $\beta << {\pi \over \kappa}$ it is
senseless to perform asymptotic expansions of the digamma
functions in (\ref{eq:msevenz})
for large values of the argument. Rather one could perform
a Taylor expansion leading to the result
\ben
{\partial \Delta \over \partial \mu}
= -{1 \over 2\pi}\Psi (\half) + {\beta^2 \over 48\pi^3}\Psi''(\half)~
[1+3\kappa^2] -{\beta^4 \over 768\pi^5}\Psi''''(\half) [\kappa^4
+2\kappa^2+{7 \over 15}] + \cdots
\label{eq:mmone}
\een

Alternatively one may work with the full expression (\ref{eq:mtwo}).
For high temperatures one may expand the factor
factor ${(\pi t/\kappa \beta) / \ksinh (\pi t/\kappa\beta)}$
in powers of $e^{-{2\pi t \over \kappa \beta}}$.
One gets
\ben
{\partial \Delta \over \partial \mu} = {1 \over \beta}\sum_{n=0}^\infty
{\rm Re}~\Psi '[\half(1+ {\pi(2n+1) \over \beta}+i\kappa]
\label{eq:mnine}
\een
For small $\beta$ we can use the
asymptotic expansions for $\Psi ' (z)$ for large
$z$ and thus obtain an expansion for ${\partial \Delta
\over \partial \mu}$ in powers of $\beta^2$. The result agrees
entirely with the expansion (\ref{eq:mmone}).

The expression for the entropy is obtained by integrating the
expression (\ref{eq:mmone}) and using
(\ref{eq:mthreea})
\ben
S = - {\beta^3 \Psi''(\half) \over 48}[2\kappa^2 + \kappa^4]
+ {\beta^5 \Psi''''(\half) \over 1440}[7\kappa^2+5\kappa^4+\kappa^6] +\cdots
\label{eq:meleven}
\een
Note that $\Psi''(\half)$ and $\Psi''''(\half)$ are negative so that
the leading contribution to the entropy is positive.

One important feature of the high temperature limit is that
there is no term proportional to the ``volume'', which is
$\klog~\mu$ in this model.

Contrary to the results of the genus expansion the entropy has a
regular behaviour at high temperatures. However the specific
heat $C_v = -\beta {\partial S \over \partial \beta}$ is
negative. This means that the system is unstable. We should
not really trust the above thermodynamic expressions for all
temperatures.

We believe that this instability is related to an inherent
nonperturbative instability of the model defined naively
as an inverted harmonic oscillator.
In fact it is quite
unclear how should one define the matrix model so that it
satisfies all the basic physical requirements \cite{POLC}.
In a properly defined model there should be no instability of
the kind discussed above. One possible scenario could be  :
the specific heat should remain positive, but the entropy
saturates at high temperatures.
It is also possible that there is a phase transition in the
singlet sector itself at $\beta \sim {1 \over \kappa}$. Such
a phase transition is distinct from the usual KT transition in
this model which is driven by the non-singlet states
and which is a {\em perturbative} phenomenon.

What is clear from the above discussion, however, is that
the genus expansion is a bad guide to the behaviour of the
thermodynamic entropy at high temperatures and that nonperturbative
effects are of crucial importance in a discussion of the geometric
entropy.

Even if we obtain a physically reasonable expression for the
thermodynamic entropy it is still not clear how one could use
this result to obtain the geometric entropy. This is because
in this theory interactions are not translationally invariant
and it is far from obvious how one could extract an entropy density
which we expect to be position dependent as well. From the point
of view of the direct calculation of the geometric entropy
discussed in the earlier part of this paper this issue is related
to the fact that the entropy would depend on the region of space
which is integrated out to obtain the density matrix.

To obtain the geometric entropy of the underlying string theory,
one further needs to address the question of the exact correspondence
of the collective field and the massless scalar of the string theory.
The point is, the massless scalar of the string theory seems to be
related to the collective field in a {\em non-local} (in space) way
\cite{POLC}.
Hence the geometric entropy obtained by integrating out the string
theory scalar field in some region of the liouville space is very
different from that obtained by integrating out the fermions or
the collective fields in some region of the $x$ or $\tau$ space.
Nevertheless our results strongly indicate that in all these
quantities nonperturbative stronng coupling effects play a crucial
role.

Finally to really understand black hole entropy in this string
theory one needs a clear understanding of the black hole
solution in the matrix model. There have been several suggestions
about this \cite{MMBH}, but the situation is rather unclear. In particular
there is no contact with known black hole solutions of the low energy
effective field theory with nontrivial tachyon backgrounds
\cite{KRAMA}. All
these issues are intimately tied with a deeper understanding of
the nature of space of time in the matrix model.

\noindent {\bf Acknowledgements}: I would like to thank
S. Govindarajan, G. Mandal, A. Sen, S. Shenker, S. Wadia
and especially S. Mathur for discussions. I would like to
thank A. Jevicki for many discussions and collaboration
during the initial stages of this work.

\end{document}